\documentclass{webofc}

\usepackage[varg]{txfonts}   
\usepackage{hyperref}
\usepackage{url}
\hypersetup{colorlinks=true,citecolor=blue,urlcolor=blue,linkcolor=blue}

\begin{document}

\title{Baryon Electric Charge Correlation as QCD Magnetometer}

\author{\firstname{Heng-Tong} \lastname{Ding}\inst{1} \and
        \firstname{Jin-Biao} \lastname{Gu}\inst{1}\and
        \firstname{Arpith} \lastname{Kumar}\inst{1}\fnsep\thanks{\email{arpithk@ccnu.edu.cn}} \and
        \firstname{Sheng-Tai} \lastname{Li}\inst{1}
}

\institute{Key Laboratory of Quark and Lepton Physics (MOE) and Institute of
Particle Physics,
     Central China Normal University, Wuhan 430079, China }

\abstract{
The detection of strong magnetic fields in peripheral heavy-ion collisions is crucial for observing effects such as the chiral magnetic effect but has proven exceptionally difficult. To address this, we propose the baryon electric charge correlation $\chi^{\rm BQ}_{11}$ and the chemical potential ratio $\mu_{\rm Q}/\mu_{\rm B}$ as sensitive probes of magnetic fields, based on (2+1)-flavor lattice QCD simulations at the physical pion mass. Along the transition line, $\chi^{\rm BQ}_{11}$ and $(\mu_{\rm Q}/\mu_{\rm B})_{\rm LO}$ in Pb--Pb collisions increase by factors of 2.1 and 2.4 at $eB \simeq 8M_\pi^2$, respectively.  To bridge theoretical predictions with experimental observables, we implement systematic kinematic cuts that emulate detector acceptances of the STAR and ALICE experiments within the hadron resonance gas model. This allows us to construct experimentally relevant proxy observables. Furthermore, we demonstrate that $(\mu_{\rm Q}/\mu_{\rm B})_{\rm LO}$ is also sensitive to the collision system, showing a $1.5$-fold increase from Zr--Zr to Ru--Ru isobar collisions. Our findings offer new insights into thermo-magnetic effects and provide experimentally relevant guidance for the detection of magnetic fields in heavy-ion collisions.}

\maketitle

\section{Introduction}
\label{sec:intro}

Strong magnetic fields produced in off-central heavy-ion collisions have attracted considerable interest recently~\cite{Kharzeev:2007jp}. Model estimates suggest early-stage strengths $eB\sim5~M_\pi^2$ at the RHIC and $eB\sim70~M_\pi^2$ at the LHC for Au/Pb nuclei collisions~\cite{Deng:2012pc,Skokov:2009qp}. Although transient, 
a sufficiently sustained lifetime—supported by the medium's electrical conductivity and paramagnetic properties—could allow these fields to induce notable non-perturbative effects on the produced QGP. This opens the possibility for their detection through final-state observables.

Fluctuations of and correlations among net baryon number (B), electric charge (Q), and strangeness (S) have long served as---both theoretically and experimentally accessible---powerful tools for probing the QCD phase structure~\cite{HotQCD:2012fhj,STAR:2019ans, ALICE:2025mkk}. 
However, theoretical studies of these fluctuations in a magnetic background remain scarce, mostly confined to effective models~\cite{Adhikari:2024bfa}. First-principles lattice QCD calculations are essential to establish model-independent benchmarks. Initial studies with a larger-than-physical pion mass ($M_\pi\simeq 220~\rm MeV$) at a single lattice spacing~\cite{Ding:2021cwv} have recently been extended to physical pion mass~\cite{Ding:2023bft,Ding:2025jfz}. Furthermore, these fluctuations of conserved charges are now being utilized to construct the equation of state in a magnetic background and nonzero density~\cite{Borsanyi:2023buy,Astrakhantsev:2024mat, Kumar:2025ikm, Ding:2025nyh}.

In this proceedings, we present lattice QCD computations for baryon electric charge correlation and electric charge over baryon chemical potential in nonzero magnetic field with physical pions~\cite{Ding:2023bft,Ding:2025jfz}.
To bridge these theoretical results with experiment, we employ the Hadron Resonance Gas (HRG) model to construct experimentally feasible proxy observables and define systematic kinematic cuts that account for detector acceptance limitations.

\section{Fluctuations of conserved charges and HRG framework}
\label{sec:HRG}
In magnetized thermal medium, fluctuations and correlations of conserved charges $(\rm B, Q, S)$ follow from derivatives of pressure $(P)$ to chemical potential $(\hat{\mu}_X\equiv{\mu}_X/T)$ at vanishing values:
\begin{equation}
    \chi_{i j k}^{\mathrm{BQS}}=\frac{\partial^{i+j+k} }{\partial\hat{\mu}_{\mathrm{B}} ^i \partial\hat{\mu}_{\mathrm{Q}} ^j \partial\hat{\mu}_{\mathrm{S}} ^k} \left({P}/{T^4}\right)~\Bigg|_{~\hat{\mu}_{\mathrm{B}, \mathrm{Q}, \mathrm{S}}=0} \quad \text{for} \qquad P=(T/V)\ln\mathcal{Z}(V,T,\hat{\mu}_{\rm B,Q,S},eB)\,.
\end{equation}
Here, $\{i,j,k\} \in \mathrm{Z}$, with $i+j+k=2$ for second (leading)-order, grand canonical partition function $\mathcal{Z}$ and magnetic field strength $eB$. We compute continuum estimates of these fluctuations using lattice QCD simulations on $32^3\times 8$ and $48^3\times 12$ lattices with HISQ action at physical pion mass, focusing on temperatures around $T_{pc}$ and magnetic strengths up to $\sim 8M_{\pi}^2$.

Within HRG framework, external magnetic fields (along $y$-direction) significantly modify the transverse momentum phase space of charged hadrons, $\int \mathrm{d}^3 \boldsymbol{p} = \int \left|q_R\right| B~ \mathrm{d} l ~\mathrm{d} p_y ~\mathrm{d} \phi_p$. Thereby, individual resonance pressure contributions can be written as~\cite{Ding:2021cwv,Ding:2023bft,Ding:2025jfz}: 
\begin{equation}
    \frac{P^c_{R, ~\rm HRG}}{T^4}=\frac{\left|q_R\right| B}{(2 \pi)^3 T^3} \sum_{s_y
    =-s_R}^{s_R} \sum_{l=0}^{\infty} \int_{0}^{\infty}  \mathrm{d} p_y \int_{0}^{2\pi}  \mathrm{d} \phi_p   ~ \sum_{k=1}^{\infty} {(\pm 1)^{k+1}} \frac{e^{-k\left(E^c_R-\mu_R\right)/T}}{k} ,
    \label{eq:pcR}
\end{equation}
for Landau quantized levels $E^c_R(p_y,l,s_y)=\sqrt{m_R^2+ p_y^2+2\left|q_R\right| B(l+1/2-s_y)}$ with $\mu_R=\mu_{\rm B} {\rm B}_R+ \mu_{\rm Q}{\rm Q}_R+\mu_{\rm S}{\rm S}_R$. 

In experiments, these fluctuations are accessed through final-state hadrons: protons ($p$), pions ($\pi$), and kaons ($K$). To align with experiments, we construct proxies using net-conserved charges $\text{net-\{B,Q,S\}} \to ~\{\tilde{p},~{Q}^{\rm PID}\equiv \tilde{\pi}^+ + \tilde{p}+\tilde{K}^+,~\tilde{K}^+\}$~\cite{Ding:2023bft,Ding:2025jfz}: $\sigma_{p,Q^{\rm PID},K}^{i,j,k}=\sum_R (\omega_{R\rightarrow \tilde{p}})^i~ (\omega_{R\rightarrow {Q}^{\rm PID}})^j ~\left(\omega_{R\rightarrow \tilde{K}}\right)^k \times I_2^{R} $, where $I_2^{R} = {\partial^2 (P_R / T^4)}/{\partial \hat{\mu}_R^2}~\big|_{\hat{\mu}_{\mathrm{B}, \mathrm{Q}, \mathrm{S}}=0}$. To reflect experimental conditions, we restrict the momentum space integration with a Heaviside step function $\Theta$ that imposes cuts on transverse momentum and pseudo-rapidity based on detector acceptance $\left( \int \mathrm{d} p_y \int \mathrm{d} \phi_p  \right)\times ~ \Theta(p_{T_{\min}} p_{T_{\max}}, \eta_{\min}, \eta_{\max})$, 
introducing cuts as $ \omega^{cuts}_{\pi,K,p}  = I_2^{R\in\{{\pi,K,p} \},~cuts}/I_2^{R\in\{{\pi,K,p} \}} $, such that $\sigma_{p,Q^{\rm PID},K}^{i,j,k} :\omega_{R\rightarrow~\tilde{p},\tilde{\pi},\tilde{K}}~~~ \longrightarrow ~~~ \sigma_{p,Q^{\rm PID},K}^{i,j,k;~ cuts} :\omega_{R\rightarrow~\tilde{p},\tilde{\pi},\tilde{K}}~\omega^{ cuts}_{p,\pi,K} $~\cite{Ding:2025jfz}.

\section{Results}
\label{sec:results_chiBQ}

\begin{figure}[h]
\centering
\includegraphics[width=6cm,clip]{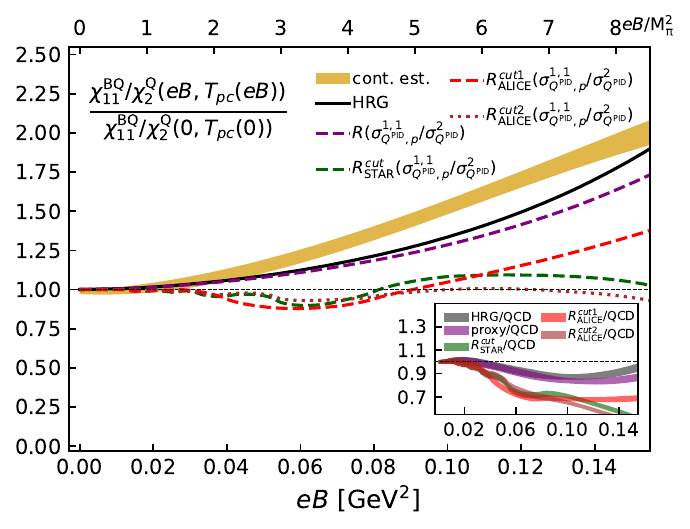}
\includegraphics[width=6cm,clip]{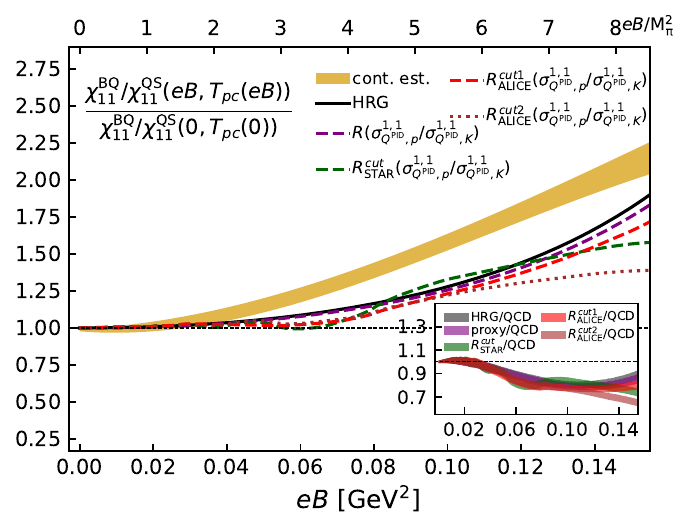}
\caption{Double ratios  $R\left(\chi^{\rm BQ}_{11} /\chi^{\rm Q}_{2}\right)$ (left), $R\left(\chi^{\rm BQ}_{11} /\chi^{\rm QS}_{11}\right)$ (right) along transition line $T_{pc}(eB)$. Bands represent lattice continuum estimates, while solid and broken lines represent results within HRG model. The figure is taken from Ref.~\cite{Ding:2025jfz}.}
\label{fig-BQ-Tpc}       
\end{figure}

Recent lattice QCD results have demonstrated a striking sensitivity of baryon electric charge correlations, $\chi^{\rm BQ}_{11}\equiv\chi^{\rm BQ}_{11}(T,eB)$, to magnetic field strengths~\cite{Ding:2023bft,Ding:2025jfz}. In heavy-ion collisions, the magnetic field effects are expected to vary with centrality classes. To this extent, we propose $\chi^{\rm BQ}_{11}$-based $R_{cp}$-like (central-peripheral) observables, $R(\mathcal{O}) \equiv \mathcal{O}\left( eB,T_{pc}(eB)\right)~\big/ ~\mathcal{O}\left(eB=0,T_{pc}(0) \right)$, shown for $\mathcal{O} \in \{\chi^{\rm BQ}_{11} /\chi^{\rm Q}_{2},~~\chi^{\rm BQ}_{11} /\chi^{\rm QS}_{11}\}$ in~\autoref{fig-BQ-Tpc}. Lattice continuum estimates exhibit significant enhancements of $\sim2$ for $\chi^{\rm BQ}_{11} /\chi^{\rm Q}_{2}$ (left) and an even more pronounced $\sim2.25$ for $\chi^{\rm BQ}_{11} /\chi^{\rm QS}_{11}$ (right) at $eB\simeq8~{M_{\pi}^2}$. Such remarkable enhancements establish potential utility of $\chi^{\rm BQ}_{11}$ as a magnetometer in QCD. Note that such double-ratio observables are well-suited for experiments to focus on $eB$-induced enhancements and mitigate volume-dependent effects~\cite{STAR:2019ans,ALICE:2025mkk}. 

Within the HRG framework, for experimental feasibility,~\autoref{fig-BQ-Tpc} also shows the corresponding proxies $R( \sigma_{Q^{\rm PID},p}^{1,1} ~\big/~\sigma_{Q^{\rm PID}}^{2}  )$ (left) and $R( \sigma_{Q^{\rm PID},p}^{1,1} ~\big/~\sigma_{Q^{\rm PID},K}^{1,1}  )$ (right), and their respective
kinematic cut results emulating STAR and ALICE detector acceptances $R^{cut}_{\rm ALICE/STAR}$. These proxies retain at least $\sim80\%$ of the magnetic sensitivity predicted by lattice QCD, as highlighted in the inset. Incorporating kinematic cuts into proxies exhibits enhancements up to 25\% at $eB\simeq8~{M_{\pi}^2}$ for $\chi^{\rm BQ}_{11} /\chi^{\rm Q}_{2}$, while strikingly up to $60\%$ for $\chi^{\rm BQ}_{11} /\chi^{\rm QS}_{11}$. This underscores their potential to isolate magnetic field signatures in heavy-ion collisions. Thus, by bridging theoretical predictions with detector-level analyses, the HRG-based proxies offer a pathway for probing magnetic effects in QCD matter through accessible fluctuation observables. 
These predictions are now being tested experimentally~\cite{Nonaka:2023xkg,ALICE:2025mkk}. Alongside ongoing efforts, the ALICE collaboration has already reported centrality-dependent enhancements in the double ratio $\chi^{\rm BQ}_{11}/\chi^{\rm Q}_{2}$~\cite{ALICE:2025mkk}, a finding which is qualitatively consistent with our theoretical results. We also emphasize the ratio $\chi^{\rm BQ}_{11}/\chi^{\rm QS}_{11}$ as a promising experimental observable. Although it has not yet been measured, our results predict it to be significantly more sensitive.

\begin{figure}
\centering
\sidecaption
\includegraphics[width=6cm,clip]{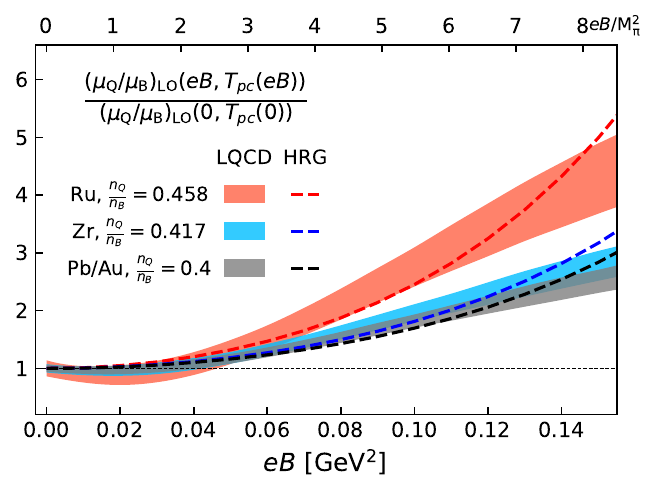}
\caption{Ratio observable $R(\left(\mu_{\rm Q}/\mu_{\rm B} \right)_{\rm LO})$ along the transition line $T_{pc}(eB)$ for several collision systems. Bands represent lattice continuum estimates for Pb/Au (grey), Ru (red), and Zr (cyan) systems, while dashed lines indicate the corresponding HRG results. The figure is taken from Ref.~\cite{Ding:2023bft}.}
\label{fig-muQ-muB}       
\end{figure}

\autoref{fig-muQ-muB} presents lattice results for $R_{cp}$-like ratio of electric charge to baryon chemical potential, $R(\left(\mu_{\rm Q}/\mu_{\rm B} \right)_{\rm LO})$, calculated along transition line for various isospin parameters, $n_{\rm Q}/n_{\rm B}$, which correspond to different collision systems. In Pb/Au collisions, $R(\left(\mu_{\rm Q}/\mu_{\rm B} \right)_{\rm LO})$ reaches approximately 2.4 at $eB\simeq8~{M_{\pi}^2}$, with a similar enhancement observed for the isobar system with Zr nuclei. In contrast, the slightly more isospin-symmetric Ru system exhibits a significantly steeper enhancement, reaching about $4$ at  $eB\simeq8~{M_{\pi}^2}$--- reflecting a roughly $1.5$-fold stronger magnetic sensitivity. Furthermore, results obtained from the HRG model (denoted by the broken lines) show reasonably good agreement with the lattice QCD data. This consistency supports the feasibility of extracting $eB$-dependencies of $\mu_{\rm Q}/\mu_{\rm B}$ by fitting particle yields within an HRG framework that incorporates a magnetized hadron spectrum.

\bibliography{refs} 

\begin{thebibliography}{15}

\bibitem{Kharzeev:2007jp}
D.E. Kharzeev, L.D. McLerran, H.J. Warringa, {The Effects of topological charge change in heavy ion collisions: 'Event by event P and CP violation'}, Nucl. Phys. A \textbf{803}, 227 (2008), \texttt{0711.0950}. \doiwoc{10.1016/j.nuclphysa.2008.02.298}

\bibitem{Deng:2012pc}
W.T. Deng, X.G. Huang, {Event-by-event generation of electromagnetic fields in heavy-ion collisions}, Phys. Rev. C \textbf{85}, 044907 (2012), \texttt{1201.5108}. \doiwoc{10.1103/PhysRevC.85.044907}

\bibitem{Skokov:2009qp}
V.~Skokov, A.Y. Illarionov, V.~Toneev, {Estimate of the magnetic field strength in heavy-ion collisions}, Int. J. Mod. Phys. A \textbf{24}, 5925 (2009), \texttt{0907.1396}. \doiwoc{10.1142/S0217751X09047570}

\bibitem{HotQCD:2012fhj}
A.~Bazavov et~al. (HotQCD), {Fluctuations and Correlations of net baryon number, electric charge, and strangeness: A comparison of lattice QCD results with the hadron resonance gas model}, Phys. Rev. D \textbf{86}, 034509 (2012), \texttt{1203.0784}. \doiwoc{10.1103/PhysRevD.86.034509}

\bibitem{STAR:2019ans}
J.~Adam et~al. (STAR), {Collision-energy dependence of second-order off-diagonal and diagonal cumulants of net-charge, net-proton, and net-kaon multiplicity distributions in Au + Au collisions}, Phys. Rev. C \textbf{100}, 014902 (2019), [Erratum: Phys.Rev.C 105, 029901 (2022)], \texttt{1903.05370}. \doiwoc{10.1103/PhysRevC.100.014902}

\bibitem{ALICE:2025mkk}
S.~Acharya et~al. (ALICE), {Measurement of correlations among net-charge, net-proton, and net-kaon multiplicity distributions in Pb-Pb collisions at $\sqrt{s_\text{NN}}=5.02$ TeV} (2025), \texttt{2503.18743}.

\bibitem{Adhikari:2024bfa}
P.~Adhikari et~al., {Strongly interacting matter in extreme magnetic fields} (2024), \texttt{2412.18632}.

\bibitem{Ding:2021cwv}
H.T. Ding, S.T. Li, Q.~Shi, X.D. Wang, {Fluctuations and correlations of net baryon number, electric charge and strangeness in a background magnetic field}, Eur. Phys. J. A \textbf{57}, 202 (2021), \texttt{2104.06843}. \doiwoc{10.1140/epja/s10050-021-00519-3}

\bibitem{Ding:2023bft}
H.T. Ding, J.B. Gu, A.~Kumar, S.T. Li, J.H. Liu, {Baryon Electric Charge Correlation as a Magnetometer of QCD}, Phys. Rev. Lett. \textbf{132}, 201903 (2024), \texttt{2312.08860}. \doiwoc{10.1103/PhysRevLett.132.201903}

\bibitem{Ding:2025jfz}
H.T. Ding, J.B. Gu, A.~Kumar, S.T. Li, {Second order fluctuations of conserved charges in external magnetic fields}, Phys. Rev. D \textbf{111}, 114522 (2025), \texttt{2503.18467}. \doiwoc{10.1103/tgm5-jvyf}

\bibitem{Borsanyi:2023buy}
S.~Borsanyi, B.~Brandt, G.~Endrodi, J.~N.~Guenther, R.~Kara, A.D. Marques~Valois, {QCD equation of state in the presence of magnetic fields at low density}, PoS \textbf{LATTICE2023}, 164 (2024), \texttt{2312.15118}. \doiwoc{10.22323/1.453.0164}

\bibitem{Astrakhantsev:2024mat}
N.~Astrakhantsev, V.V. Braguta, A.Y. Kotov, A.A. Roenko, {QCD equation of state at nonzero baryon density in an external magnetic field}, Phys. Rev. D \textbf{109}, 094511 (2024), \texttt{2403.07783}. \doiwoc{10.1103/PhysRevD.109.094511}

\bibitem{Kumar:2025ikm}
A.~Kumar, H.T. Ding, J.B. Gu, S.T. Li, {QCD Equation of State with Strong Magnetic Fields and Nonzero Baryon Density}, PoS \textbf{LATTICE2024}, 175 (2025), \texttt{2502.03152}. \doiwoc{10.22323/1.466.0175}

\bibitem{Ding:2025nyh}
H.T. Ding, J.B. Gu, A.~Kumar, S.T. Li, {Leading-Order QCD Equation of State in Strong Magnetic Fields at Nonzero Baryon Chemical Potential} (2025), \texttt{2508.07532}.

\bibitem{Nonaka:2023xkg}
T.~Nonaka, {Experimental Overview on Fluctuations of Conserved Charges}, Acta Phys. Polon. Supp. \textbf{16}, 1 (2023). \doiwoc{10.5506/APhysPolBSupp.16.1-A14}

\end{thebibliography}
\end{document}